\documentclass[apj]{emulateapj}
\usepackage{amssymb, amsmath, apjfonts, natbib, color, hyperref}

\hypersetup{
    colorlinks=true,
    linkcolor=blue,
    citecolor=blue,
}

\newdimen\digitwidth
\setbox1=\hbox{0}
\digitwidth=\wd1
\catcode`"=\active
\def"{\kern\digitwidth}

\slugcomment{ApJ Accepted}

\begin{document}

% The title
\title{Metallicity Properties of the Galactic Bulge Stars Near and Far: Expectations from the Auriga Simulation}

% The author
\author{Bin-Hui Chen \altaffilmark{1, 2} and Zhao-Yu Li\altaffilmark{1, 2, 3}}
\altaffiltext{1}{Department of Astronomy, School of Physics and Astronomy, Shanghai Jiao Tong University, 800 Dongchuan Road, Shanghai 200240, China}
\altaffiltext{2}{Key Laboratory for Particle Astrophysics and Cosmology (MOE) / Shanghai Key Laboratory for Particle Physics and Cosmology, Shanghai 200240, China}
\altaffiltext{3}{Correspondence should be addressed to: lizy.astro@sjtu.edu.cn}

% The abstract
\begin{abstract}
Using the high resolution Milky Way-like model from Auriga simulation we study the chemical properties of the Galactic bulge, focusing on the metallicity difference between stars on the near side (in front of the Galactic center) and the far side (behind the Galactic center). In general, along certain sight lines the near side is more metal-rich than the far side, consistent with the negative vertical metallicity gradient of the disk, since the far side is located higher above the disk plane than the near side. However, at the region $l<0^\circ$ and $|b|\le6^\circ$, the near side is even more metal-poor than the far side, and their difference changes with the Galactic longitude. This is mainly due to the fact that stars around the minor axis of the bar are more metal-poor than those around the major axis. Since the bar is tilted, in the negative longitude region, the near side is mainly contributed by stars close to the minor axis region than the far side to result in such metallicity difference. We extract stars in the X-shape structure by identifying the overdensities in the near and far sides. Their metallicity properties are consistent with the results of the whole Galactic bulge. The boxy/peanut-shaped bulge can naturally explain the metallicity difference of the double red clump stars in observation. There is no need to involve a classical bulge component with different stellar populations.
\end{abstract}


% Introduction----------------------------------------------------------------------------------
\section{Introduction}
	 The Galactic bulge holds important clues of the formation history of Milky Way (MW) \citep{barbuy_etal_2018}. It has a mixture of stellar populations with different chemical properties \citep{zoccal_etal_2003,zoccal_etal_2008,zoccal_etal_2010,ness_etal_2012,ness_etal_2013,ness_etal_2013_2,ness_etal_2014,zoccal_etal_2014,barbuy_etal_2018}. The stars in the Galactic bulge are generally old and metal-rich \citep{ortola_etal_1995, kui_ric_2002, ferrer_etal_2003, sahu_etal_2006, clarks_etal_2008, brown_etal_2009, brown_etal_2010,clarks_etal_2011,valent_etal_2013, calami_etal_2014, gennar_etal_2015} with a negative vertical metallicity gradient \citep{zoccal_etal_2008, babusi_etal_2010, hill_etal_2011, bensby_etal_2013, ness_etal_2013, gonzal_etal_2013, rojas_etal_2014, rojas_etal_2017}. The observed parallelogram shaped morphology reflects the existence of an almost end-on bar with a boxy/peanut-shaped  structure, that is produced by the buckling instability of the bar \citep{raha_etal_1991, mer_sel_1994,mihos_etal_1995, bur_ath_2005, debatt_etal_2006, martin_etal_2006} or other much slower process (see \citealt{sel_ger_2020} and reference there in). There are also pieces of evidence for a possible classical bulge embedded in the bar region \citep{barbuy_etal_2018, zoccal_etal_2018}, which could be less significant as suggested by \citet{shen_etal_2010}.
	 
	 Observations of the red clump (RC) stars revealed an X-shaped structure in the Galactic bulge \citep{mcw_zoc_2010, nataf_etal_2010} which was later associated with the boxy/peanut-shaped  bar after the buckling instability \citep{li_she_2012, ness_etal_2012, li_she_2015}. The X-shaped structure reveals itself as a bimodal distribution of the apparent magnitude of the red clump stars, which are commonly used as standard candles to trace the bulge structure. \citet{gonzal_etal_2015} summarized the main observational properties of the RC double peaks:
	 
	 1.$\;$The double peaks are clearly visible in $|b| \gtrsim 5^{\circ}$ region and they merge to a single peak in $|b| \lesssim 5^{\circ}$ (P1).
	 
	 2.$\;$The magnitude difference between the two peaks increases with the increasing absolute value of the Galactic latitude (P2).
	 
	 3. At a fixed Galactic latitude, the brighter RC is more dominant in the positive Galactic longitude region than the fainter one, which is more important in negative longitudes. In $|l| > 4^\circ$ region only one of them is visible (P3).
	 
	 4. The variation of the RC magnitudes as well as the magnitude difference between the two RCs are independent with the observational bands (P4).

	 All these properties are consistent with an X-shaped structure in the bulge (X-shape scenario). \citet{li_she_2015} also pointed out that there is no simple letter ``X''-like structure in the Galactic bulge, but a boxy/peanut-shaped component; the X-shape is likely a visual effect caused by the concave curved isodensity contours in the inner region of the peanut bulge.
	 
     \begin{figure}[htbp!]
    \centering
    \includegraphics[width=.45\textwidth, height=.445\textheight]{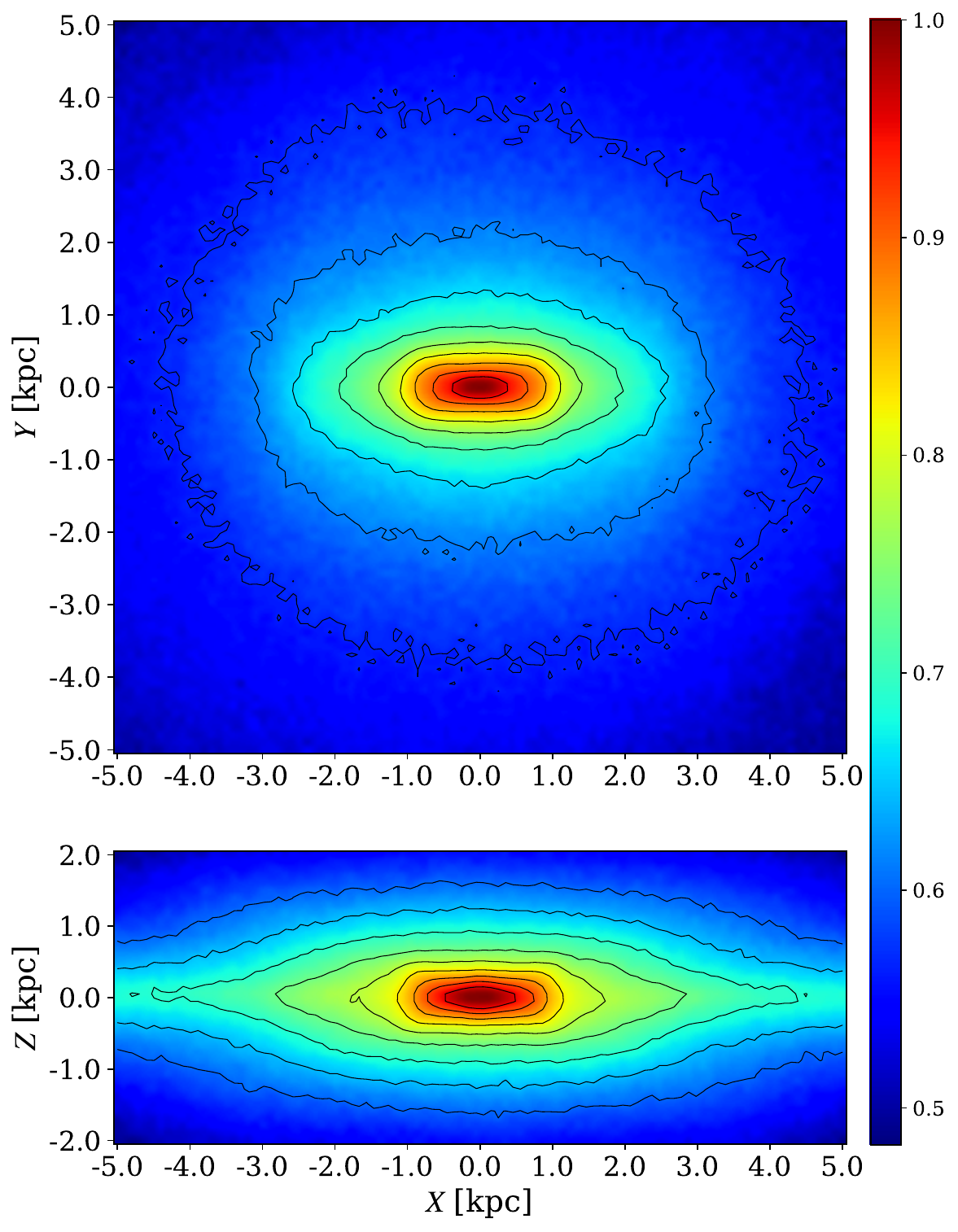}
    \caption{The face-on and edge-on density maps of the stellar disk of AU23 with logarithmic density contours overplotted. All the maps are normalized with respect to their maximal number density. The semi-major axis of the bar is about 3 kpc.}
    \label{Fig1}
    \end{figure}

	 Recently, \citet{lee_etal_2015} and \citet{joo_etal_2017} proposed a different scenario to use the multiple population classical bulge to explain the split of RC stars (MCB scenario). They argued that the magnitude difference between the two RCs reflects the intrinsic luminosity difference of two stellar populations in a classical bulge, with an additional unbuckled bar embedded in; the intrinsic magnitude difference of the two RC populations originates from the helium abundance difference between the first generation (G1) stars and later generations (G2+). Such phenomena have been observed in globular clusters (see \citealt{lee_etal_2015} and references there in). In \citet{lee_etal_2015}, their Fig.~4 showed a superposition of the classical bulge and a thin bar in the MCB scenario, with the two different RC populations in the classical bulge only. The bar component blurs the magnitude difference of the double RCs. Under this scenario, only one peak is observable near the mid-plane of the MW (P1). With larger $|b|$, such blurring effect becomes weaker to make the double peaks gradually visible, with their magnitude difference increasing with the Galactic latitude (P2). Since the bar is tilted, the blurring effect is different in different longitudinal regions, leading to the spatial variation of the relative importance between the double peaks (P3). Condition P4 is also accounted for in this scenario.

	 Metallicity information could be a key factor in discriminating the two scenarios. The MCB scenario predicts an intrinsic metallicity difference between the two RCs' stars \citep{lee_etal_2015,joo_etal_2017}, which was supported by following observations \citep{lee_etal_2019,dongw_etal_2021}. However, \citet{gonzal_etal_2015} argued that the X-shape scenario could also produce a similar but smaller metallicity difference between the two RCs. In the X-shape scenario, the MW has a vertically thickened boxy/peanut-shaped bar, indicating that the Galaxy evolution is dominated by the secular process. On the other hand, the MCB scenario has a dominant classical bulge component corresponding to a more violent early merger history. Therefore, towards the Galactic bulge region, in a given line-of-sight, the double peaks in the apparent magnitude distribution will have different metallicity properties in the two scenarios. In the X-shape scenario at intermediate Galactic latitudes, stars are bar dominated so the metallicity difference between the bright (near side) and faint (far side) of the RC stars in the Galactic bulge is expected to vary in different ($l,\,\,b$). For MCB scenario, a dominant classical bulge component will result in similar metallicity difference between the bright and faint RC stars across the Galactic bulge region.
	 
	 Aiming to better understand the Galactic bulge structure and chemical properties, with the high resolution MW-like N-body model Auriga halo 23 (AU23) in \citet{grand_etal_2017}, we investigate chemical properties of the Galactic bulge stars, particularly focusing on the difference between stars in the near and far sides of the Galactic center. In Section 2, we briefly describe the model used here. In Section 3, we explore the metallicity properties of the Galactic bulge of AU23. In Section 4, we compare the simulation results with observations and discuss the influence of the bar angle on the metallicity difference between the near and far sides. The results are summarized in Section 5.

    \begin{figure*}[htbp!]
    \centering
    \includegraphics[width=1.\textwidth, height=.3\textheight]{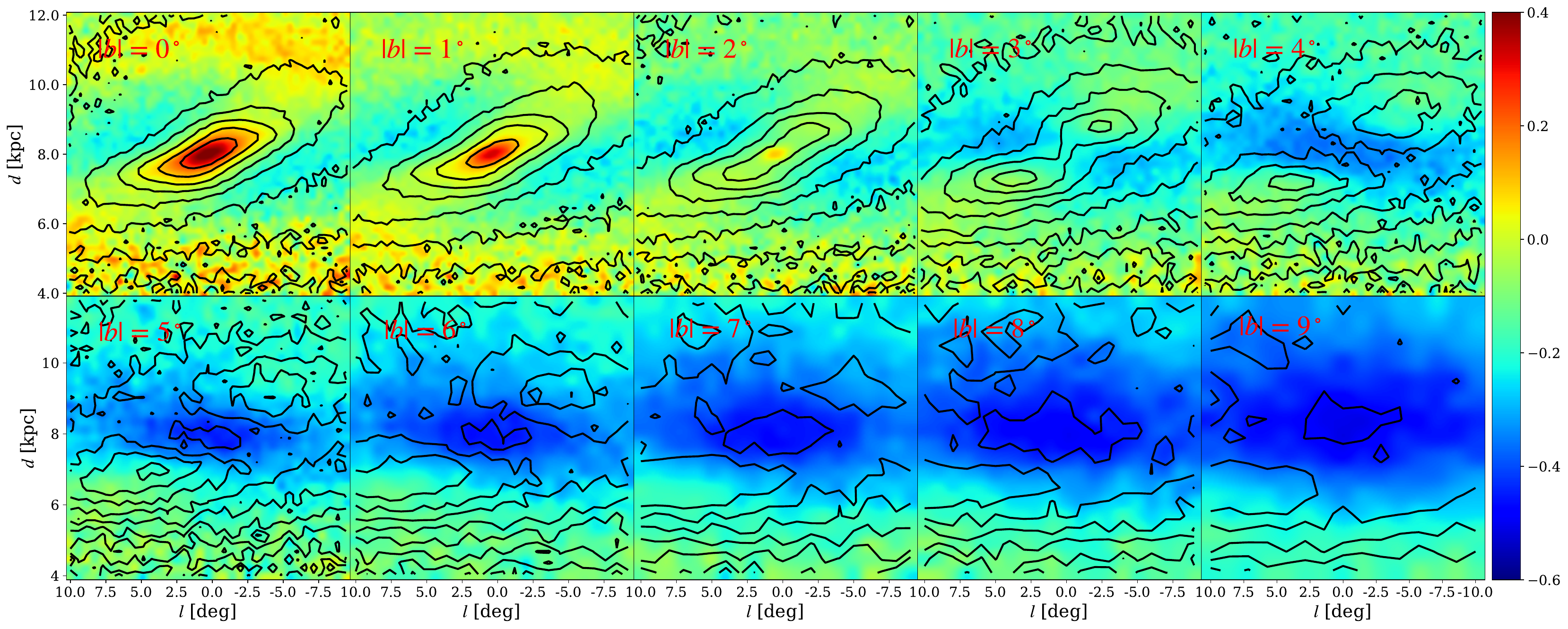}
    \caption{The longitude-distance distribution of stars in different Galactic latitude slices in AU23 color-coded with the average [Fe/H] values. The background isodensity contours show the spatial distribution of the stars younger than 11 Gyr. There are obvious double over-density regions for $3^\circ\,\leq|b|\,\leq7^\circ$ slices. The  over-density region at the far side disappears in $|b|\ge8^\circ$ slice. The double over-density regions gradually approach to each other as $|b|$ decreasing and finally merge in $|b| \le3^\circ$ slices.}
    \label{Fig2}
    \end{figure*}
    
    \begin{figure*}[htbp!]
    \centering
    \includegraphics[width=1. \textwidth, height=.3\textheight]{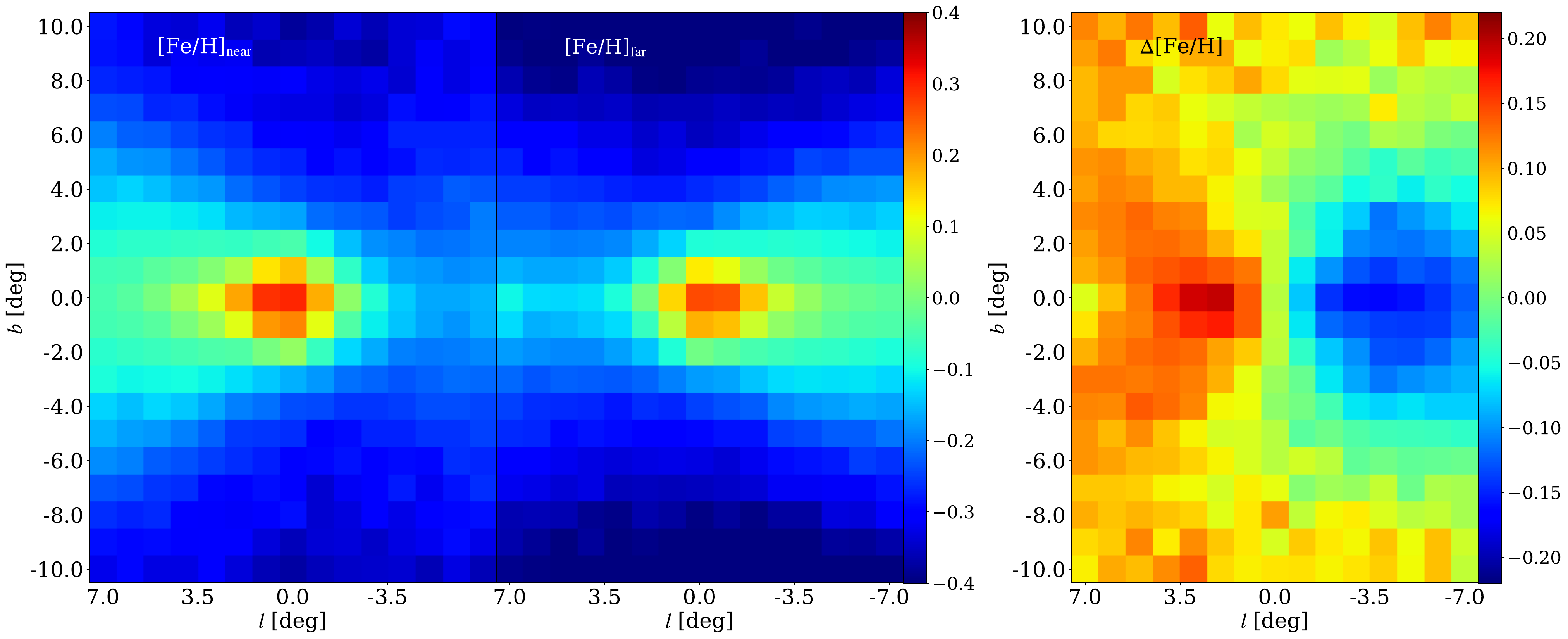}
    \caption{The left two panels show the average metallicity maps of the near side (<8 kpc) and far side (>8 kpc) of the Galactic bulge from AU23, with their difference shown in the right panel ($\Delta$[Fe/H]=[Fe/H]$_{\rm{near}}$ $-$ [Fe/H]$_{\rm{far}}$). We exclude the stars <4 kpc or >12 kpc that are not located in the bulge. At larger Galactic latitude regions ($|b|\,>\,6^\circ$), the near side is more metal-rich than the far side at all longitudes. However, closer to the mid plane($|b|\,\le\,6^\circ$), at positive Galactic longitude region the near side is more metal-rich than the far side, and vice versa in the negative longitude side.}
    \label{Fig3}
    \end{figure*}
    \section{Numerical Model of The Galactic Bulge}\label{Sec2}
    
    Auriga project introduces 30 cosmological magneto-hydrodynamical zoom-in simulations of galaxies in isolated MW-like mass dark halos \citep{grand_etal_2017}, namely AU1$-$AU30. \citet{fragko_etal_2020} systematically analyzed the chemodynamics of the simulations in the Auriga project. Among all the galaxies in the Auriga project, AU23 exhibits a clear boxy/peanut-shaped bar similar to the Galactic bulge, as shown in Fig.~2 of \citet{grand_etal_2017} and Fig.~4 of \citet{fragko_etal_2020}. Mass of the baryons (stars and gas) in AU23 is about $9\,\times\,10^{10}\,\, M_{\odot}$ in a dark matter halo of $\sim1.6\,\times\,10^{12}\,\,M_{\odot}$. The disk scale length $R_d$ is about $4.99$ kpc and the disk to total stellar mass ratio ($D/T$) is 0.63, with an accreted stellar mass fraction within $0.1 R_{200}$ at $\sim0.1$. The metallicity of the single stellar population (stellar particles in simulation) is determined by the initial metallicity of the parent gas cell, and the metal return for SN II, SN Ia and asymptotic giant branch (AGB) stars are concerned in each time-step\footnote{More details can be found in \citet{grand_etal_2017} and \citet{fragko_etal_2020}.}. In this paper, we set the solar position at ($X$, $Y$, $Z$) = (-8, 0, 0.02) kpc with the Galactic center (GC) at (0, 0, 0) kpc and the angle between the bar major axis and Sun-GC line at $30^\circ$.
	
	Fig.~\ref{Fig1} shows the face-on and edge-on density maps of the stellar disk in AU23, which exhibits a clear boxy/peanut-shaped bulge. \citet{grand_etal_2018} showed that the stars in AU23 reproduced a chemical thin/thick disk dichotomy as seen in the MW. The thick $\alpha$-rich disk was produced via a merger-induced starburst event 10-11 Gyr ago, after that the thin low-$\alpha$ disk and bar began to form. Therefore, the stars younger than 11 Gyr in AU23 are more bar-like with strong boxy/peanut-shaped bulge. The older stars older than 11 Gyr resembles a spheroidal structure similar to a possible classical bulge or inner halo structure, contributing to $\sim15\%$ of the total stellar mass. Similar tendency has been found in both observations (e.g. \citealt{babusi_etal_2010, ness_etal_2012, catchp_etal_2016}) and other simulations (e.g. \citealt{debatt_etal_2017, portai_etal_2017}). In the following analysis, stars younger than 11 Gyr are used to trace the double peak positions in the Galactic bulge region, while the full sample (including stars older than 11 Gyr) is used to estimate the metallicity difference between the two peaks.
    
    \section{Results}
    In this section, we first show the metallicity distribution of stars in each latitude slice. Then we explore the metallicity difference between the stars in the near and far sides of the bulge in details. Finally we focus on the metallicity difference between the stars in the double peak position in the Galactic bulge.
    
    \subsection{Metallicity Distributions in Latitudinal Slices of the Galactic Bulge}
    In Fig.~\ref{Fig2}, we divide the stars in AU23 into different Galactic latitude slices, from $b=-9^\circ$ to $b=9^\circ$ with $1^\circ$ height for each slice. We further superpose the opposite latitudinal slices ($\pm b$) to increase the number statistics. In each panel of Fig.~\ref{Fig2} ([Fe/H] color-coded), the contour represents the number density distribution of stars younger than 11 Gyr to highlight the position of the double peaks, useful to identify the stars belonging to the X-shape (similar to observations). In order to better visualize the density contours, larger bins are used in the slices at $|b|>6^\circ$. Including all the stars will produce a smooth distribution. In the slices at $|b|\simeq5^\circ$, there are two clear over-density regions, corresponding to the double peaks in the apparent magnitude distributions of the RC stars as in observations \citep{mcw_zoc_2010, nataf_etal_2010} and previous simulation works (e.g. \citealt{li_she_2012, ness_etal_2012}).
    
    Close to the mid-plane, such over-density regions gradually approach to each other and merger together in slices $|b|\le3^\circ$ (P1 and P2). In slices $|b|\ge4^\circ$ the double peaks are visible only in $|l|\lesssim2.5^\circ$ for $|b|=4^\circ$ and $|l|\lesssim3.5^\circ$ for $|b|=5^\circ$, and different peaks dominate different sides (P3).
    
    From the metallicity distribution in Fig.~\ref{Fig2}, we can see that in slices $|b|\ge4^\circ$, the two over-density regions are more metal-rich than the region between them. In slices $|b|\le3^\circ$, the stellar metallicity is higher close to GC and along the major axis of the bar. The slices closer to the mid-plane are more metal-rich. In slices $|b|>4^\circ$, the near side (bright RCs) seems slightly more metal-rich than the far-side (faint RCs).

    \begin{figure*}[htbp!]
    \centering
    \includegraphics[width=1.\textwidth, height=.315\textheight]{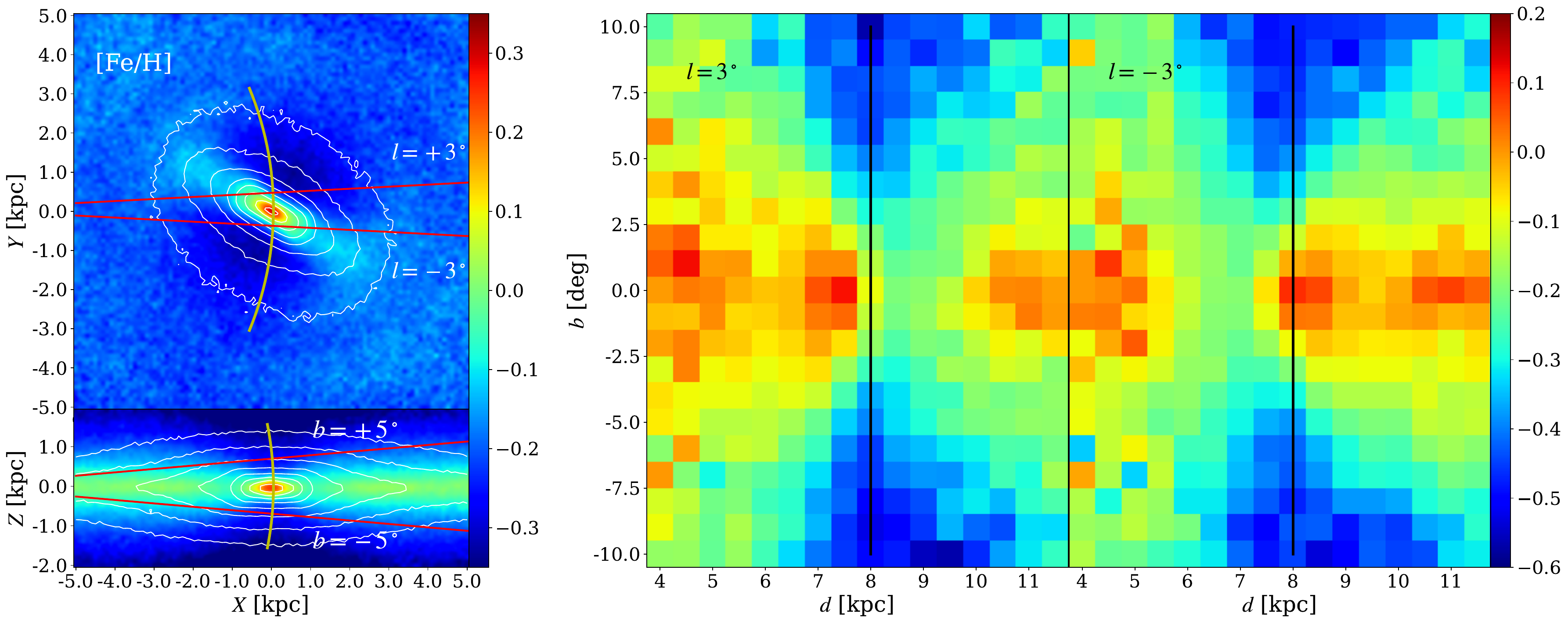}
    \caption{Left panel: Face-on (top) and edge-on (bottom) maps of the stars in AU23 color-coded with the average [Fe/H] values, with the white contours representing the density distributions. The yellow arc represents the 8 kpc distance from the Sun, which separates the near and far sides of the bulge. The two red solid lines denote different lines of sight with Galactic longitude value at $\pm3^\circ$, the metallicity maps at such directions are explicitly shown in the right two panels. Near the mid-plane, for the positive longitudinal region ($l=3^\circ$), the line of sight will go through the bar region (<8 kpc, near side) first, and then a metal-poor region at the bar minor axis direction (>8 kpc, far side), to produce a positive metallicity difference value ([Fe/H]$_{\rm near}$ $-$ [Fe/H]$_{\rm far}$ > 0). Similarly a negative metallicity difference value is produced in the negative longitudinal region. Such difference becomes smaller further away from the mid-plane.}
    \label{Fig4}
    \end{figure*}
    
    \begin{figure*}[htbp!]
    \centering
    \includegraphics[width=.425\textwidth, height=.85\textheight]{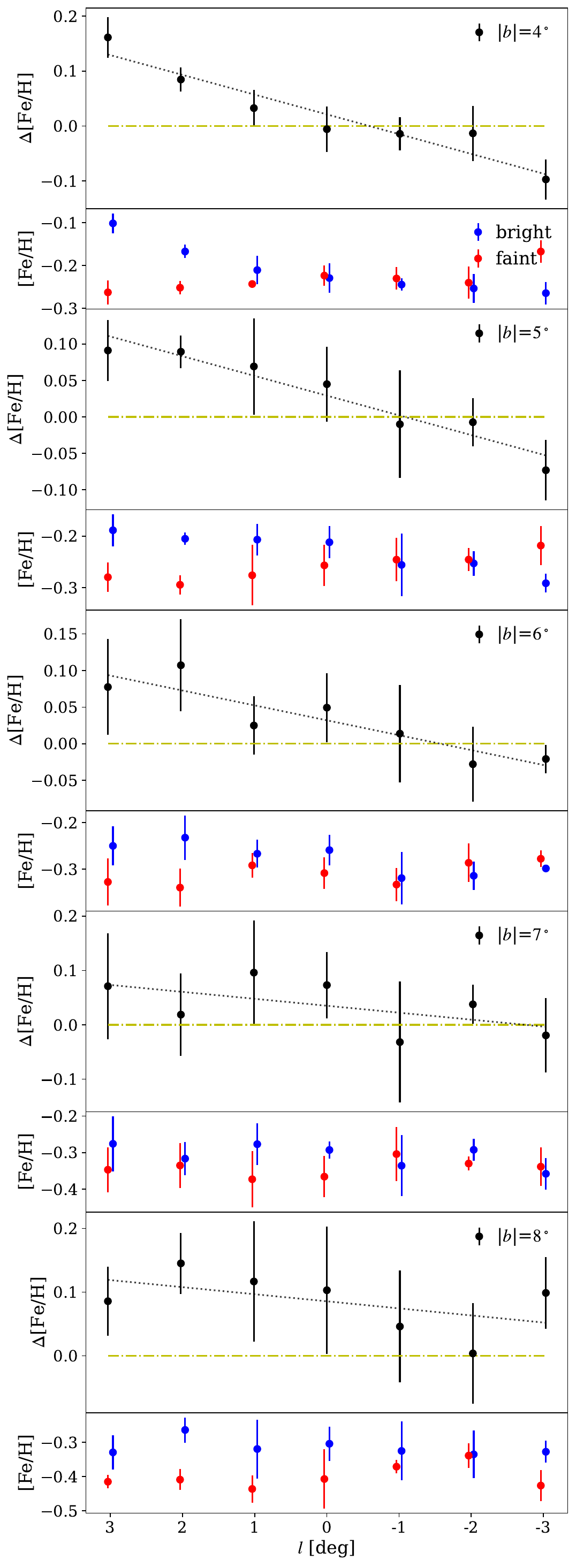}
    \includegraphics[width=.425\textwidth, height=.85\textheight]{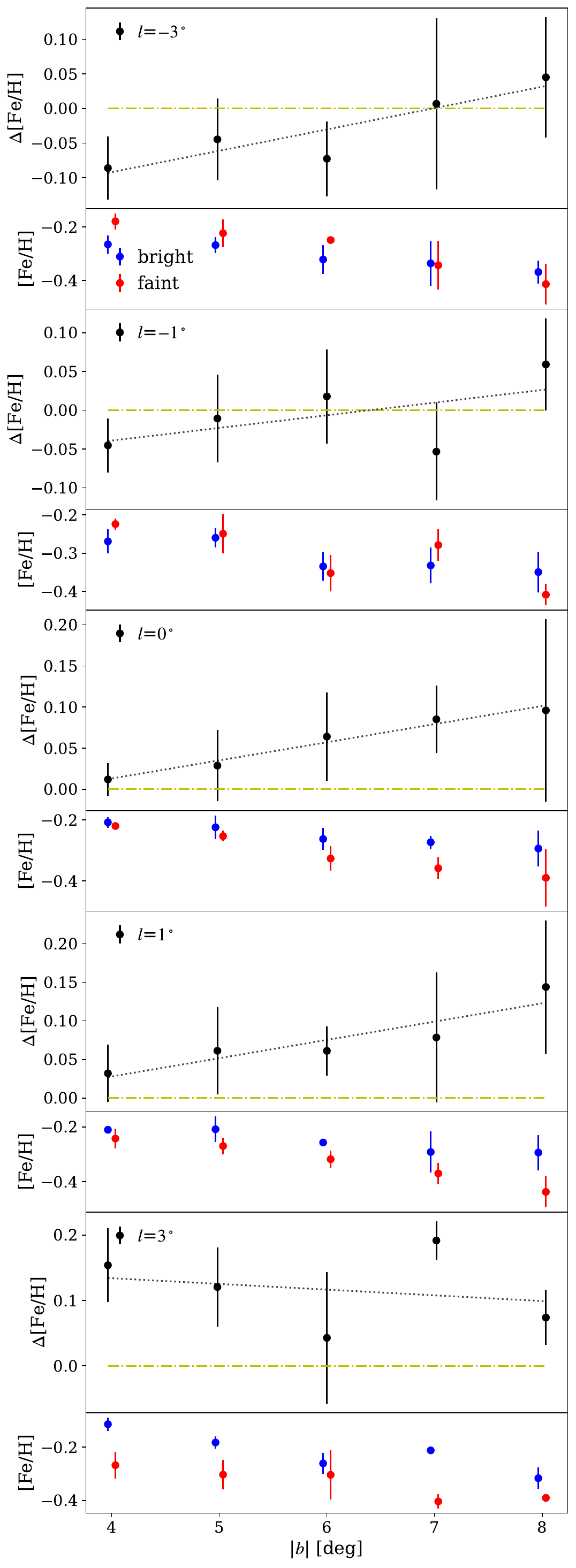}
    \caption{Left: The longitudinal profiles of [Fe/H] and $\Delta$[Fe/H] of the double peaks for five different $|b|$ slices (from $|b|=4^\circ$ to $8^\circ$). Right: The latitudinal profiles of [Fe/H] and $\Delta$[Fe/H] of the double peaks for five different $l$ slices (from $l=-3^\circ$ to $3^\circ$). The blue points are the closer/brighter peaks and red points for farther/fainter ones. The data points and error bars are the mean value and standard deviation of the [Fe/H] in each bin, respectively, with the bin size of 1 square degree. In each $\Delta$[Fe/H] panel, the black dotted line shows the linear fitting result and the yellow dot-dashed line shows the zero difference level. In the slices $|b|=4^\circ$ and $5^\circ$, $\Delta$[Fe/H] has a clear linear correlation with respect to the Galactic longitude, where the gradient is $\simeq0.03$ dex/deg by the linear fitting. For higher latitudinal regions the $\Delta$[Fe/H] values become almost positive, with a flat horizontal gradient of $\Delta$[Fe/H]. In the right column at the slice $l=-3^\circ$ and $-1^\circ$, there is a tendency that $\Delta$[Fe/H] increases from negative to positive at higher latitudes. Closing to $l=0^\circ$, $\Delta$[Fe/H] generally becomes positive. There is also clear evidence of the vertical metallicity gradient in all slices for both the near and far sides.}
    \label{Fig5}
    \end{figure*}
    
    \subsection{Metallicity Difference Between the Near and Far Sides of the Galactic Bulge}
    According to the stellar distance from the GC, we can divide the stars into two groups: 4 kpc$<d<$8 kpc as the near side, and 8 kpc$<d<$12 kpc as the far side. By analyzing the metallicity properties, we can get a qualitative picture of the metallicity difference between the near and far sides of the bulge.
    
    Fig.~\ref{Fig3} shows the [Fe/H] distribution in the $l-b$ diagram of the two populations. As shown in the left two panels, the most metal-rich region of the two populations are located at different Galactic longitude sides. The right panel shows the [Fe/H] difference of the two populations, with $\Delta$[Fe/H]=[Fe/H]$_{\rm{near}}$ $-$ [Fe/H]$_{\rm{far}}$ (the difference between previous two panels). In the high latitude region ($|b|>6^\circ$) the near side is more metal-rich than the farther one at all longitudes. This is due to the negative vertical metallicity gradient of the MW; along certain line-of-sight towards the Galactic bulge region, the fainter/farther peak is located at larger vertical height above the disk plane than brighter/closer one. In addition, in the right panel we can see a horizontal [Fe/H] gradient in the intermediate latitude region. Close to the mid-plane ($|b| \leq 6^\circ$) at negative longitude side ($l<0^\circ$), the metallicity difference even becomes negative. This is unexpected, since the vertical metallicity gradient of the disk should indicate the near side to be more metal-rich than the far side.
    
    To better understand the variation of $\Delta$[Fe/H] across the bulge region, we show the faced-on and edge-on maps of AU23 color-coded with the mean [Fe/H] in the left panel of Fig.~\ref{Fig4}. The metallicity is larger along the major axis of the bar than that along the minor axis in the face-on map. The more metal-rich (younger) stars are more concentrated around the bar major axis with higher metallicity, while the metal-poor (older) stars show rounder distribution with smaller ellipticity, as shown in Fig.~A1 of \citealt{fragko_etal_2020}. It can help to understand the phenomena in Fig.~\ref{Fig3}. The right panels in Fig.~\ref{Fig4} show the latitude-distance distributions of stars in two vertical slices at $l=\pm3^\circ$. The black solid line marks the position of the GC distance at 8 kpc. For the sight line at $l=3^\circ$, close to the mid-plane ($|b|\leq 6^\circ$), the near side is dominated by the stars in the major axis of the bar (metal-rich), while the far side is populated by the stars in the minor axis, which is more metal-poor. On the contrary, for the sight line at $l=-3^\circ$ and $|b|\leq 6^\circ$, the near side is mainly populated by stars in the minor axis region (metal-poor) and the far side is mainly dominated by the major axis region of the bar (metal-rich). This difference further enhances the horizontal metallicity gradient close to the mid-plane. Interestingly, in the left panels of Fig.~\ref{Fig4} the most metal-rich region of the bar seems to coincide with the central boxy core identified in \citet{li_she_2015} for numerical buckled bars.
    
    In the region away from the mid-plane, the metallicities of both the near and far sides decrease, with the far side showing a steeper latitudinal gradient. Therefore, at both positive and negative longitude regions, the near side gradually becomes more metal-rich than the far side. This is demonstrated in the lower panel of an edge-on projection of the model with two lines of sight at $b=\pm5^\circ$. The vertical metallicity gradient is more important to result in a positive metallicity difference between the near and far sides. At such latitude range we can still find a longitudinal gradient of the metallicity difference. The negative longitude side generally shows smaller (or even negative) metallicity difference. This difference arises from the fact that at negative longitude region, the near side is mainly contributed by the more metal-poor minor axis of the bar than the more metal-rich far side around the major axis. The latitudinal metallicity gradient decreases faster in the far side, to gradually result in a positive metallicity difference at larger latitude regions.

    \subsection{Metallicity Difference between the Bright and Faint peaks of the X-shape}
    The results in Section 3.2 hint that the bright (near) and faint (far) peaks of the X-shape may show similar patterns of the metallicity difference across the Galactic bulge region. To quantify such phenomenon more precisely, we transform the stellar spatial distribution into distance histograms for different $(l,\,\,b)$ fields, similar to Fig.~2 in \citet{li_she_2012}. To better visualize the double peaks in the distance distribution, we mainly focus on stars younger than 11 Gyr (as shown in Fig.~\ref{Fig2}) to determine the peak position. Note that the double peaks can be identified in the region $|l|<3^\circ$ and $3^\circ\lesssim|b|\lesssim8^\circ$. Then we apply a Gaussian mixture model provided by \texttt{scikit-learn} \citep{pedreg_etal_2012}, which implements an expectation-maximization algorithm, to identify the location of the double peaks. More details are shown in the Appendix. In the following analysis, we pick out stars (including all age populations) that lie within 0.25 kpc distance from the peaks to study their metallicity.
    
    The left column of Fig.~\ref{Fig5} shows the mean metallicity and metallicity differences as a function of the longitude in different latitude slices from $|b|=4^\circ$ to $8^\circ$ of the bright and faint double peaks. For lower Galactic latitude slices ($|b|=4^\circ,\,\,5^\circ,\,\,6^\circ$), with $l$ increasing, [Fe/H] of the brighter (fainter) peak gradually increases (decreases). $\Delta$[Fe/H] gradually increases from negative (at $l<0^\circ$) to positive values (at $l>0^\circ$) due to the metallicity difference between the major and minor axis of the bar demonstrated in Fig.~\ref{Fig4}. In slices $|b|\ge6^\circ$ almost all the metallicity difference values are positive due to the vertical metallicity gradient.
    
    The right column of Fig.~\ref{Fig5} shows the similar result but in different Galactic longitude slices ($l\,=\,-3^\circ,\,-1^\circ,\,0^\circ,\,1^\circ,\,3^\circ$). At each slice, the brighter peak always has a shallower latitudinal metallicity gradient than the fainter one, since at the same latitude the fainter/farther stars are located at larger vertical heights than the brighter/closer stars. In the $l=-3^\circ$ slice, $\Delta$[Fe/H] gradually increases from negative to positive values with increasing $|b|$. For $l\ge0^\circ$ slices, at different latitudes, $\Delta$[Fe/H] values are all positive. This has been demonstrated in Fig.~\ref{Fig4}: at $l\ge0^\circ$, the near side is mainly contributed by the major axis part of the bar, which is more metal-rich than the minor axis part to produce a positive metallicity difference.
    
    Therefore, according to the simulation, we can see that the near side is not always more metal-rich than the far side. The near side is even more metal-poor than the far side at $-3^\circ<l<-1^\circ$ and $|b|<6^\circ$. As we have explained in Section 3.2, it is the metallicity difference between the minor axis (metal-poor) and major axis (metal-rich) of the bar to result in negative $\Delta$[Fe/H] values in such region. In the other fields, the near side is generally more metal-rich, which is mainly due to the vertical metallicity gradient in the bulge/bar structure. Although this model may not truly reflect the chemical properties of the Galactic bulge, it is still valuable to shed light on the spatial variation of the metallicity differences between the near and far sides across the bulge region.
    
    \section{Discussion}
    \subsection{Comparison with Previous Works and Implications on the MCB Scenario}
    The metallicity difference between the double peaks in the simulation is consistent with the previous observations \citep{joo_etal_2017,lee_etal_2019}. We note that the observed value of metallicity difference between the doubel RCs in \citet{dongw_etal_2021} is $0.149\pm0.036$ dex at $(l,\,\,b)=(-1^\circ,\,\,-8.5^\circ)$ that is comparable to the same window in AU23, where $\Delta$[Fe/H]=[Fe/H]$_{\rm{near}}-$[Fe/H]$_{\rm{far}}=0.157$ dex. Our results confirm that the buckled bar can naturally explain the metallicity difference between the two RC peaks. A multi-population classical bulge component, i.e., the MCB scenario, is not necessarily needed to explain the observations.
    
    Most importantly, we find a gradient of the metallicity difference along the longitudinal direction at $|b|\le6^\circ$ in the simulation shown in Figs. \ref{Fig3} and \ref{Fig5}. Such a horizontal gradient is mainly caused by two factors. The first one is that the metallicity is higher close to the major axis of the bar, and lower around the minor axis. The tilted bar angle is the second factor. However, such a phenomenon is not expected in the MCB scenario; a prominent spheroidal classical bulge tends to produce similar $\Delta$[Fe/H] values between the double RCs at the opposite $\pm l$ regions. Particularly at $|b|=5^\circ$ region, we find a monotonically increasing pattern of $\Delta$[Fe/H] with $l$. This trend should not be expected in the MCB scenario. In addition, along the latitudinal direction, we note that the vertical metallicity gradient in brighter peaks is shallower than fainter ones. In the MCB scenario, the bright and faint peaks should have no significant difference in the vertical metallicity gradient.
    
    The metallicity difference map between the near and far sides in Fig.~\ref{Fig3} shows similar pattern to the profiles in Fig.~\ref{Fig5}. We apply linear fitting to the result in Fig.~\ref{Fig3} to get a more quantitative estimation of the longitudinal (horizontal) and latitudinal (vertical) gradients of $\Delta$[Fe/H] with error estimated by Monte Carlo simulation. The results are listed in Table 1. We can see that both directions generally show positive gradients except for the latitudinal gradients at positive $l$. The longitudinal gradient is highest at smaller latitude values ($|b|=3^\circ$) and decreases towards higher latitudes. The latitudinal gradients are much higher at the negative longitudinal side than the positive longitudinal side with slightly negative gradients.
    
    Our results show that a self-consistent buckled bar in AU23 can account for the metallicity difference between the bright and faint RCs in the Galactic bulge with a similar value to the observation at the bulge field $(l,\,\,b)=(-1^\circ,-8.5^\circ)$. Compared to the MCB scenario, a boxy/peanut bulge can naturally explain the structure, kinematics, and metallicity properties of the bulge stars. On the other hand, the MCB scenario has to include an unbuckled but very thick bar (extending to $|b|\sim8^\circ$) to produce the Galactic longitude dependence of RCs at high $|b|$ region \citep{joo_etal_2017}. Such a bar configuration is actually quite unlikely to be found in the MW or other disk galaxies. Future observations of the longitudinal/latitudinal metallicity difference gradients will be the smoking gun evidence to discriminate the MCB scenario.

    \subsection{Dependence of the Metallicity Difference on the Bar Angle}
    The bar tilting angle ($\alpha$) plays an important role in generating the metallicity difference gradient along the longtitudinal directions as shown in Figs. \ref{Fig3} and \ref{Fig4}. Different bar tilting angle should result in different $\Delta$[Fe/H] pattern. From the simulation, we choose different tilting angles of the bar to produce the metallicity difference maps in the bulge region. The longitudinal gradient of $\Delta$[Fe/H] could be used to constrain the bar tilting angle.

    \begin{figure*}[htbp!]
    \centering
    \includegraphics[width=.8\textwidth, height=.55\textheight]{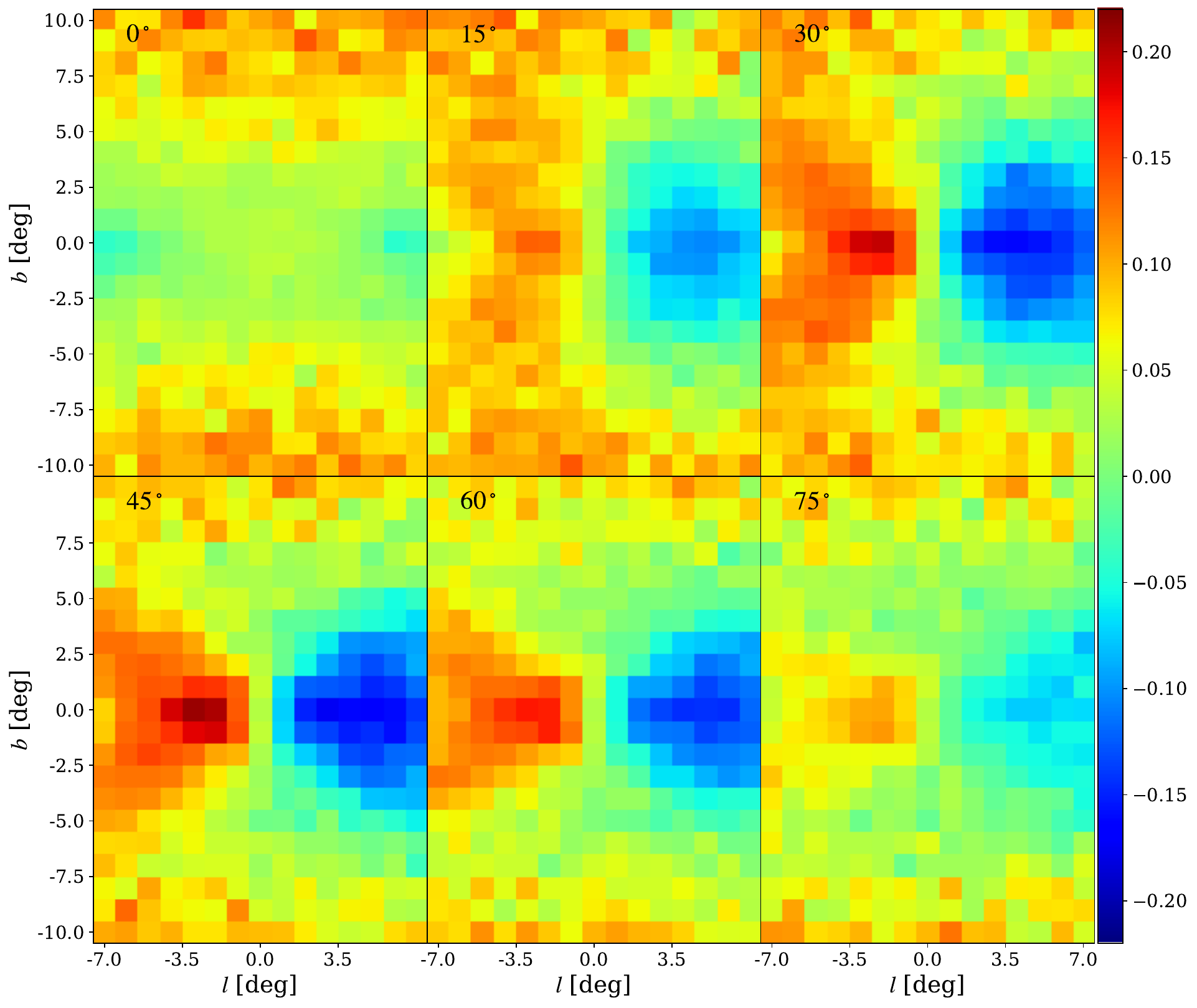}
    \caption{The bulge metallcity difference map ($\Delta$[Fe/H]) in the $l-b$ diagram at different bar angles. As the titlting angle of the bar increases, the metallicity pattern changes accordingly. In the $0^\circ$ case, viewed along the bar major axis, $\Delta$[Fe/H] is positive almost in all pixels; the distortion effect from the less metal-rich minor axis stars is negligible because in such case the line of sight always crosses the bar major axis first. The intermediate tilting angles generally show clear horizontal gradients, that is most significant at $\alpha=45^\circ$.}
    \label{Fig6}
    \end{figure*}

\begin{figure}[htbp!]
    \centering
    \includegraphics[width=.48\textwidth, height=.24\textheight]{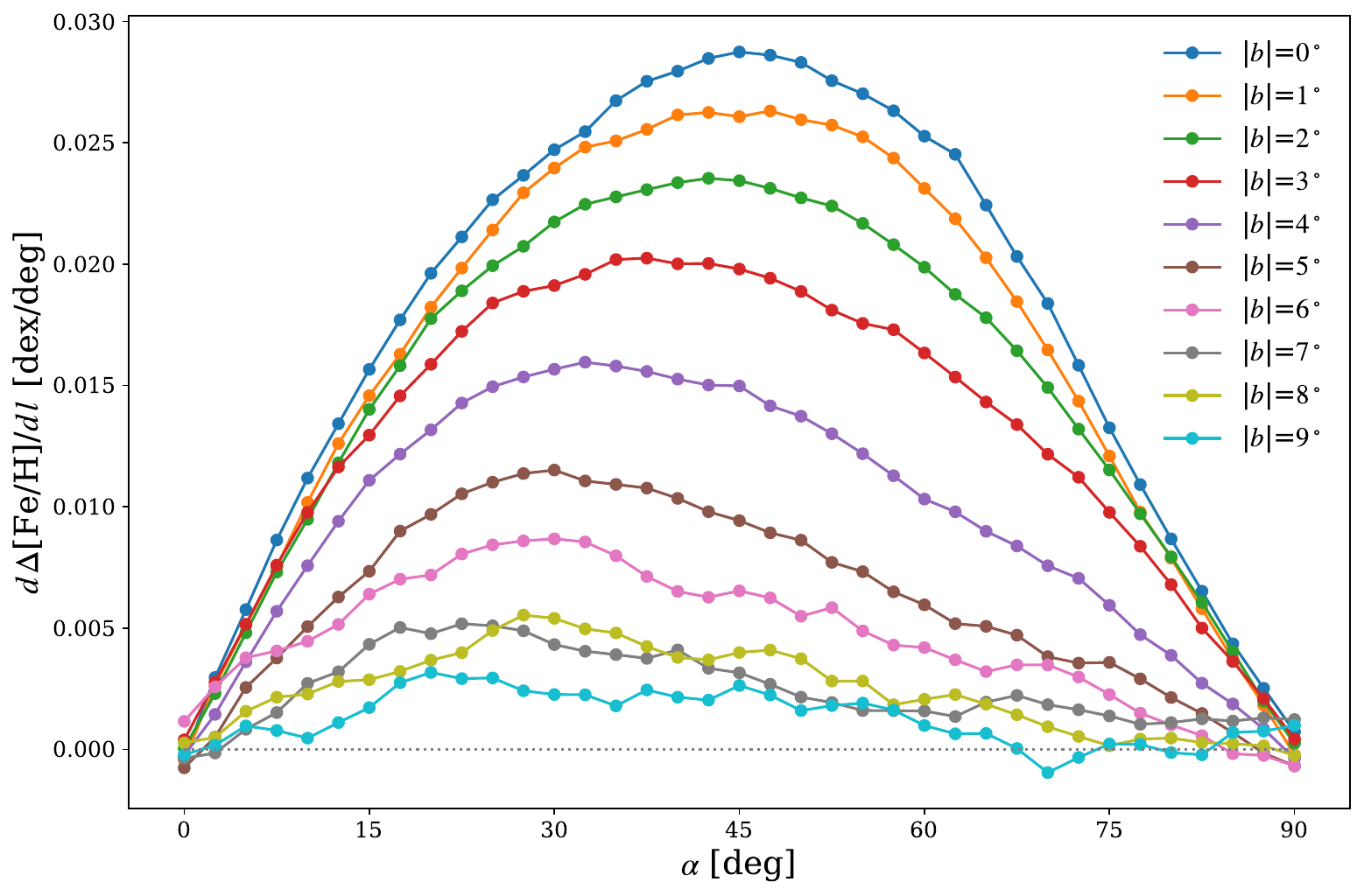}
    \caption{The longitudinal gradients of the metallicity difference $d\Delta$[Fe/H]/$d\,l$ at different $|b|$ slices in AU23 with different bar tilting angles. The gray dotted curve marks the zero level in the figure. At higher $|b|$ slices, the maximum value points are gradually smaller.}
    \label{Fig7}
    \end{figure}

    Fig.~\ref{Fig6} shows the $\Delta$[Fe/H] maps for different bar tilting angle. In the $0^\circ$ case, $\Delta$[Fe/H] values are generally positive across the bulge region at $|b|>2^\circ$. There is no horizontal gradient in $\Delta$[Fe/H]. For intermediate bar angles, there is a negative $\Delta$[Fe/H] region close to the mid-plane and a horizontal $\Delta$[Fe/H] gradient at $3^\circ<|b|<9^\circ$. The gradient $d\Delta$[Fe/H]/$d\,l$ seems sensitive to the bar tilting angle. Fig.~\ref{Fig7} shows the gradient $d\Delta$[Fe/H]/$d\,l$ at different $|b|$ slices for different $\alpha$. The gradient curves at $|b|<6^\circ$ show a sinusoidal pattern, with the steepest gradient at $\alpha\sim45^\circ$. For higher $|b|$ slices ($|b|\geq7^\circ$), the gradient is much shallower and varies mildly at different bar tilting angles. 
    
    With the gradient curves in Fig.~\ref{Fig7}, future observations of the bulge stars could help to constrain the bar tilting angle. It is suggested to measure the gradients in multiple latitudinal slices lower than $6^\circ$ since the other curves at higher latitudes in Fig.~\ref{Fig7} are relatively flat with low amplitudes ($<0.005~{\rm dex/deg}$). If the gradient in only one latitudinal slice is used, even a small error will result in a large uncertainty of the bar tilting angle. As listed in Table 1, the gradient at $|b|=5^\circ$ is 0.012 dex/deg with the error of 0.002, corresponding to $\alpha=15^\circ\sim45^\circ$. Therefore, to better constrain the bar angle, the error of the gradients should be small, and measurements at multiple latitudinal slices are also needed.
    
    \begin{table*}[htbp!]%调节图片位置，h：浮动；t：顶部；b:底部；p：当前位置
	\centering
	\caption{Longitudinal and Latitudinal Gradients of $\Delta[Fe/H]$ in the Bulge Region}
	\label{tab:1}  
	\begin{tabular}{cccc ccc}%表格中的数据居中，c的个数为表格的列数
		\hline\hline\noalign{\smallskip}
		$|b|$ & $3^\circ$ & $5^\circ$ & $7^\circ$ & $9^\circ$\\
		\noalign{\smallskip}\hline\noalign{\smallskip}
		$d\Delta$[Fe/H]/$d\,l$ & 0.019$\pm$0.007 & 0.012$\pm$0.002 & 0.004$\pm$0.002 & 0.002$\pm$0.002\\
		\hline\hline\noalign{\smallskip}	
		$l$ & $-5^\circ$ & $-3^\circ$ & $0^\circ$ & $3^\circ$ & $5^\circ$\\
		\noalign{\smallskip}\hline\noalign{\smallskip}
		$d\Delta$[Fe/H]/$d\,{|b|}$ & 0.026$\pm$0.002 & 0.025$\pm$0.004 & 0.005$\pm$0.003 & -0.007$\pm$0.009 & -0.003$\pm$0.002\\
		\noalign{\smallskip}\hline
	\end{tabular}
    \end{table*}
    
    \section{Conclusion}
    We analyze the Auriga simulation AU23 to investigate the chemical properties of the boxy/peanut-shaped bulge, especially focusing on the difference between the near and far sides of the Galactic bulge. By identifying the peak positions of the stellar distance distributions towards the bulge region, we can also utilize the metallicity difference properties to discriminate the X-shape scenario and the MCB scenario for the origin of the RC double peaks in the Galactic bulge.
    
    Stars in the bulge are separated into the near side ($d<8$ kpc) and far side ($d>8$ kpc) based on their relative position with respect to the Galactic center. Generally speaking, the near side is more metal-rich than the far side, especially for regions at larger latitudes. However, we find an unexpected result that the far side could be more metal-rich than near side at $l<0^\circ$ and $|b|\lesssim6^\circ$. Such metallicity difference pattern is resulted from the tilted bar and the fact that the bar major-axis region is more metal-rich than the minor-axis region. Such phenomenon can also be found for stars in the double peaks of the X-shaped structure. For the negative longitude region ($l<0^\circ$), at intermediate latitude ($|b|\le6^\circ$), the line-of-sight will first go across the metal-poor region at the near side ($d<8$ kpc) and then the metal-rich major axis region at the far side ($d>8$ kpc), resulting in the negative metallicity difference. At positive longitude region, the line of sight crosses the more metal-rich major axis for the near side and metal-poor minor axis for the far side to generate positive $\Delta$[Fe/H] values, resulting in the longitudinal $\Delta$[Fe/H] gradient in the intermediate latitude region. Because the far side is farther away from the mid-plane than the near side, the metallicity of the far side decreases faster with latitude compared to the near side. Therefore, for the negative longitude region the near side gradually becomes more metal-rich than the far side at high latitude region ($|b|\gtrsim7^\circ$), and it makes the longitudinal gradient shallower than the lower latitudinal region. This simulation may not be the real portrait of the Milky Way. However, the chemical patterns revealed in the simulation are still valuable to help us understand the evolution and formation history of the Galactic bulge.
    
    We found that a secular evolution produced boxy/peanut-shaped bulge can account for the metallicity difference between the double RCs in observations. The MCB scenario is not necessary for such phenomenon. Further more, we note that the horizontal gradient is sensitive to the bar titling angle. Such relationship can help to constrain the bar tilting angle of the MW with future observations in the Galactic bulge (e.g., SDSS-V, BDBS, Gaia, CSST, etc).
    
    \section{Acknowledgements}
    We thank the referee for helpful comments and suggestions. We also thank Robert Grand for providing the Auriga simulation, and Juntai Shen for helpful discussions. This work was supported by the National Natural Science Foundation of China under grant No. 12122301, by a Shanghai Natural Science Research Grant (21ZR1430600), by the Cultivation Project for LAMOST Scientific Payoff and Research Achievement of CAMS-CAS, by the ``111'' project of the Ministry of Education under grant No. B20019, and by the China Manned Space Project with No. CMS-CSST-2021-B03. This work made use of the Gravity Supercomputer at the Department of Astronomy, Shanghai Jiao Tong University.
    
% References ------------------------------------------------------------------------
%\bibliography{main}
%\bibliographystyle{apj}

\appendix
\section{Identifying the Double Peaks}
\begin{figure*}[htbp!]
    \centering
    \includegraphics[width=.9\textwidth, height=.8\textheight]{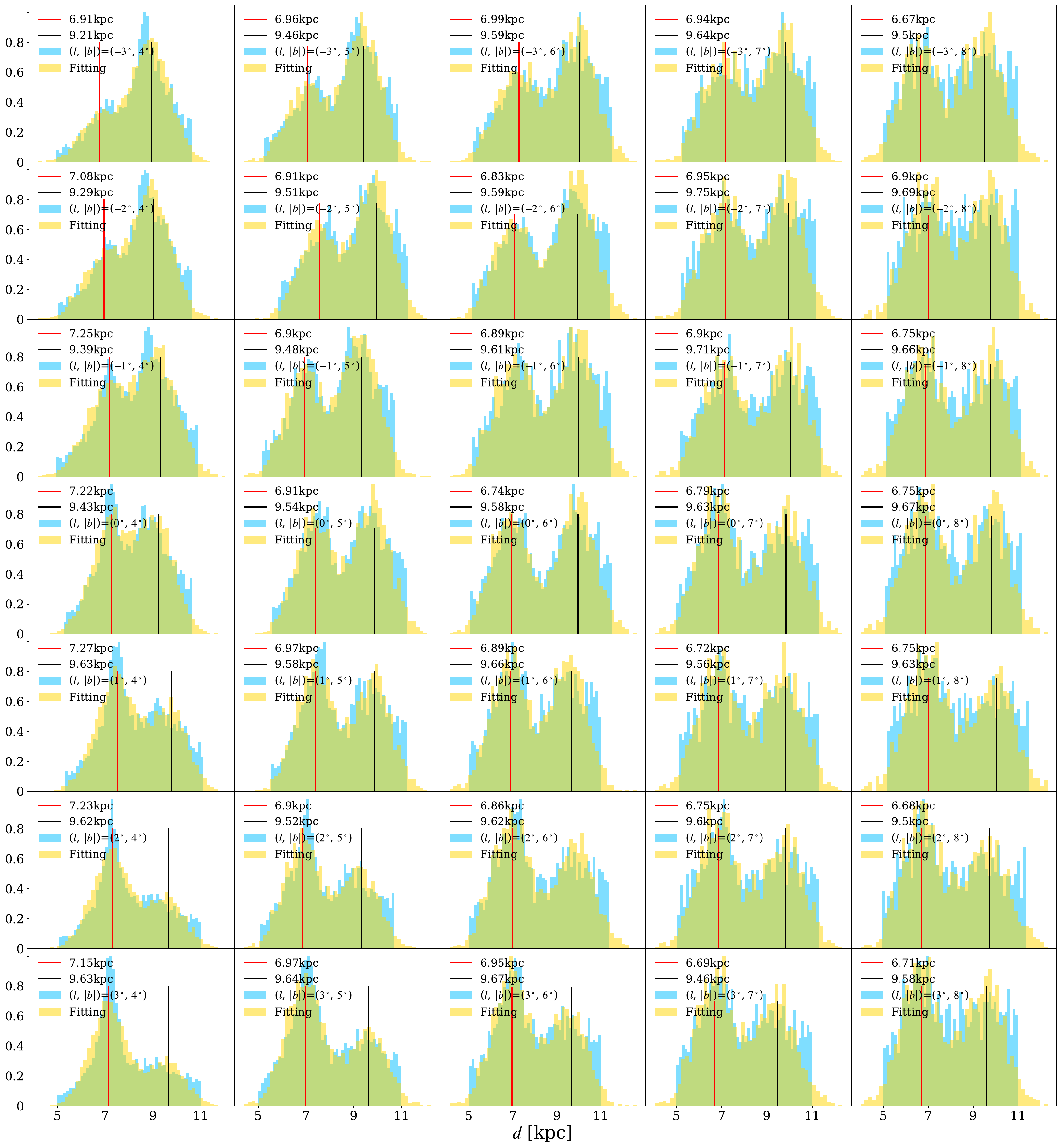}
    \caption{The Gaussian Mixture Model fitting results for the stars $<$11Gyr and $5\,{\rm kpc}<d<11\,{\rm kpc}$, where $d$ is the distance from the solar position. The blue histograms are the distance histograms at different $(l,\,\,b)$ windows that correspond the data points in Fig.~\ref{Fig5}. The yellow ones are the samples produced by the fitting results. The positions of the double peaks are shown in the upper left corner in each panel, which are marked by the red and black solid lines in the histograms.}
    \label{Fig8}
    \end{figure*}

In order to identify the double peaks in difference $(l,\,\,b)$ windows, we apply the \texttt{sklearn.mixture.GaussianMixture} provided by \texttt{scikit-learn} \citep{pedreg_etal_2012}. The results are shown in Fig.~\ref{Fig8}. The cyan histograms are the stellar distance distributions for stars at $5\,{\rm kpc}<d<11\,{\rm kpc}$ in AU23, which serve as the input for \texttt{sklearn.mixture.GaussianMixture}. The yellow ones are fitting results produced by \texttt{sample} method of \texttt{sklearn.mixture.GaussianMixture}. They are overall consistent with the peak position of the observation. The position of the closer (farther) peaks are marked by the red (black) solid lines. We use the double peaks positions in Fig.~\ref{Fig8} to find the two over-density regions in the bulge region corresponding to the X-shape structure.
\end{document}